\documentclass[11pt]{article}
\usepackage{graphicx}
\def\beq{\begin{eqnarray}}
\def\eeq{\end{eqnarray}}

\begin{document}

\title{A method for classical and quantum mechanics}

\author{Paolo Amore\footnote{paolo@ucol.mx} \\
Facultad de Ciencias, Universidad de Colima,\\
Bernal D\'{i}az del Castillo 340, \\ Colima, Colima,\\ M\'exico.}

\maketitle

\begin{abstract}
In many physical problems it is not possible to find an exact solution. However, when some parameter in the problem is small, 
one can obtain an approximate solution by expanding in this parameter. This is the basis of perturbative methods, which have 
been applied and developed practically in all areas of Physics. Unfortunately many interesting problems in Physics are of 
non-perturbative nature and it is not possible to gain insight on these problems only on the basis of perturbation theory: 
as a matter of fact it often happens that the perturbative series are not even convergent. 

In this paper we will describe a method which allows to obtain arbitrarily precise analytical approximations for the 
period of a classical oscillator. The same method is then also applied to obtain an analytical approximation to the spectrum 
of a quantum anharmonic potential by using it with the WKB method. In all these cases we observe exponential rates of convergence
to the exact solutions. An application of the method to obtain a fastly convergent series for the Riemann zeta function is also discussed.

\end{abstract}

\section{Introduction}
\label{intro}
In this article we review a method for the evaluation of a certain class of integrals 
which occurr in many physical problems. The method that we propose has been used to obtain arbitrarily precise approximations to the period 
of a classical oscillator, to the deflection angle of light by the sun and to the precession of the perihelion of a planet 
in General Relativity~\cite{AS04,AAFS04,AF04}, to the spectrum 
of a quantum potential~\cite{AL04} and to certain mathematical functions, such as the Riemann zeta function~\cite{Am04}.
This paper is organized in three sections: in section \ref{method} we outline the method and explain its general features; in section \ref{appli}
we discuss different applications of the method and present numerical results; finally, in section \ref{conclu} we  draw our conclusions.

\section{The method}
\label{method}
We consider the problem of calculating integrals of the form:
\beq
{\cal I}_{\nu} = \int_{x_-}^{x_+} \left[ F - f(x) \right]^\nu \ g(x) \ dx
\label{eq_1_1}
\eeq
where $F = f(x_\pm)$ and $f(x) \leq F$ for $x_- \leq x \leq x_+$. We also ask that $\nu > -1$ so that the singularities are integrable.
Integrals of this kind occurr for example in the evaluation of the period of a classical oscillator or in the application of the WKB
method in quantum mechanics. We wish to obtain an analytical approximation to ${\cal I}_{\nu}$ with arbitrary precision.

The idea behind the method that we propose is quite simple: we introduce a function $f_0(x)$, which depends on one or more arbitrary 
parameters (which we will call $\lambda$) and define $F_0 \equiv f_0(x_\pm)$. Although the form of $f_0(x)$ can be chosen almost arbitrarily, 
we ask that the integral of eq.~(\ref{eq_1_1}) with $F \rightarrow F_0$ and $f(x) \rightarrow f_0(x)$ can be done analytically.

In the spirit of the Linear Delta Expansion (LDE)~\cite{lde} we interpolate the original integral as follows:
\beq
{\cal I}_{\nu}^{(\delta)} = \int_{x_-}^{x_+} \left[ F_0 -f_0(x) + \delta (F -F_0 - f(x)+f_0(x)) \right]^\nu \ g(x) \ dx \ .
 \label{eq_1_2}
\eeq
This equation reduces to eq.~(\ref{eq_1_1}) in the limit $\delta = 1$, however it yields a much simpler integral when $\delta = 0$.
We therefore write eq.~(\ref{eq_1_2}) as:
\beq
{\cal I}_{\nu}^{(\delta)} = \int_{x_-}^{x_+} \left[ F_0 -f_0(x)\right]^\nu \ \left[ 1 + \delta \Delta(x) \right]^\nu \ g(x) \ dx \ .
 \label{eq_1_3}
\eeq
where we have defined
\beq
\Delta(x) \equiv \frac{F -F_0 - f(x)+f_0(x)}{F_0 -f_0(x)} \ .
 \label{eq_1_4}
\eeq

We can use the expansion
\beq
\left( 1+ x\right)^\nu = \sum_{n=0}^\infty \frac{\Gamma(\nu+1)}{\Gamma(\nu-n+1)} \ \frac{x^n}{n!}
 \label{eq_1_5}
\eeq
which converges uniformly for $|x| < 1$. 

As a result we can substitute in eq.~(\ref{eq_1_3}) the series expansion of eq.~(\ref{eq_1_5}) provided that the constraint 
$|\Delta(x) | < 1$ is met for any $x_- \leq x \leq x_+$. In general, as we will see in the next Section, this inequality 
provides restrictions on the values that the arbitrary parameter $\lambda$ can take. 
Under these conditions the integral can be substituted with a family of series (each corresponding to a different $\lambda$):
\beq
{\cal I}_{\nu}^{(\delta)} = \sum_{n=0}^\infty  \frac{\Gamma(\nu+1)}{\Gamma(\nu-n+1)\ n!} \ {\ell}_{\nu n} \delta^n
 \label{eq_1_6}
\eeq
where
\beq
{\ell}_{\nu n} \equiv \int_{x_-}^{x_+} \left[ F_0 -f_0(x)\right]^\nu \ \left[\Delta(x) \right]^n \ g(x) \ dx \ .
\eeq

We assume that {\sl each of the integrals defining ${\ell}_{\nu n}$ can be evaluated analytically}. Although we have not yet specified the
form of $f_0(x)$, which indeed will have to be chosen case by case, we already know that, if all the conditions that we have imposed
above are met we have a family of series all converging to the exact value of the integral ${\cal I}_{\nu}$, after setting $\delta = 1$.
Since the rate of convergence of the series will clearly depend on the parameter $\lambda$, we can pick the series among all the infinite 
series representing the same integral which converges faster. In fact, although $\lambda$ is a
completely arbitrary parameter, which was inserted ``ad hoc'' in the integral, and therefore the final result {\sl cannot} depend upon it.
When the series is truncated to a given finite order, we will observe a residual dependence upon $\lambda$. We invoke the Principle
of Minimal Sensitivity (PMS)~\cite{Ste81} to minimize, at least locally, such spurious dependence and thus obtain the optimal series 
representation of the integral:
\beq
\frac{\partial \ {\cal I}_{\nu}^{(N)} }{\partial \lambda}  = 0 \ ,
 \label{eq_1_7}
\eeq
where we have defined ${\cal I}_{\nu}^{(N)}$ as the series of eq.~(\ref{eq_1_6}) truncated at $n = N$ and taking $\delta = 1$.
We will see in the next Section that this simple procedure allows to obtain series representation which converge fastly.
Interestingly, in general the optimal series obtained in this way display an exponential rate of convergence.

\section{Applications}
\label{appli}

In this Section we consider different applications of the method described above. 

\subsection{The period of a classical oscillator}

As a first application, we now consider the problem of calculating the period of a unit mass 
moving in a potential $V(x)$\cite{AS04,AAFS04,AF04}. The total energy $E = \frac{\dot{x}^2}{2} + V(x)$ 
is conserved during the motion.
The exact period of the oscillations is easily obtained in terms of the integral:
\begin{equation}
T = \int_{x_-}^{x_+} \frac{\sqrt{2}}{\sqrt{E-V(x)}}dx,
\label{eq_2_1}
\end{equation}
where $x_\pm$ are the inversion points, obtained by solving the equation $E = V(x_{\pm})$.

Clearly, the integral of eq.~(\ref{eq_2_1}) is a special case of the integral considered in the 
previous section, corresponding to choosing $F = E$, $f(x) = V(x)$, $g(x)=1$ and $\nu = -1/2$.
In order to test our method we consider the Duffing  oscillator, which corresponds to the potential 
$V(x) = \frac{1}{2} \ x^2 + \frac{\mu}{4} \ x^4$.
We choose the interpolating potential to be $V_0(x) = \frac{1+\lambda^2}{2} \ x^2$ and obtain
\begin{equation}
\Delta(x) =  \frac{2}{1+\lambda^2} \ 
\left[ \frac{\mu}{4} \ (A^2+x^2) - \frac{\lambda^2}{2} \right] \ . 
\label{eq_2_2}
\end{equation}
The series in eq.~(\ref{eq_1_6}) converges to the exact period for 
$\lambda > \lambda_0 \equiv \sqrt{\frac{\mu A^2}{2}} \ \sqrt{1-\frac{1}{\mu A^2}}$,
since $|\Delta(x)| <1$ uniformly for such values of $\lambda$ and $|x|\leq A$. 

The period of the Duffing oscillator calculated to first order using 
(\ref{eq_1_6}) is then
\begin{equation}
T^{(0)}_\delta + \delta \ T^{(1)}_\delta =
\frac{2 \pi}{\sqrt{1+\lambda^2}} 
\left\{ 1 - \frac{\delta}{1+\lambda^2} \ \left[ \frac{3}{8}  \mu 
A^2-\frac{\lambda^2}{2}\right] 
\right\}
\label{eq_2_3}
\end{equation}

By setting $\delta=1$ and applying the PMS we obtain the optimal value of $\lambda$, 
$\lambda_{PMS} = \frac{\sqrt{3 \mu}}{2} A$, which remarkably coincides with the one
obtained in \cite{AA1:03} by using the LPLDE method to third order.
The period corresponding to the optimal $\lambda$ is 
\begin{equation}
T_{PMS} = \frac{4  \pi}{\sqrt{4 + 3  \mu  A^2}} ,
\label{eq_2_4}
\end{equation}
and it provides an error less than $2.2  \%$ to the exact period for any value of $\mu$ and $A$.
This remarkable result is sufficient to illustrate
the nonperturbative nature of the method that we are proposing: in fact, a perturbative approach, which
would rely on the expansion of some small {\sl natural} parameter, such as $\mu$, would only provide 
a polynomial in the parameter itself: therefore it would never be possible to reproduce the correct 
asymptotic behavior of the period in this way.

\begin{figure}
\begin{center}
\includegraphics[width=9cm]{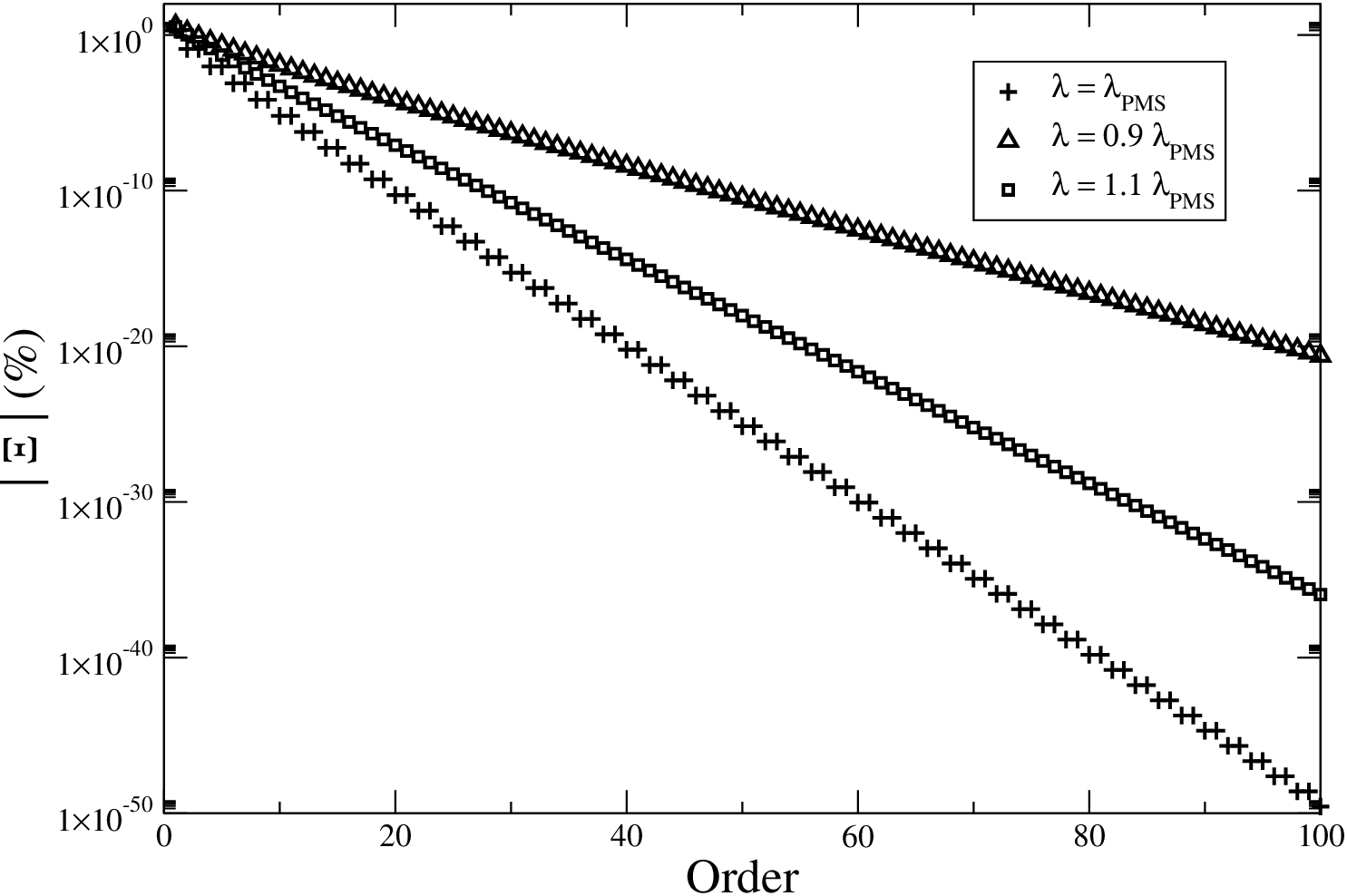}
\caption{Error over the period (in absolute value), defined as 
$\Xi \equiv \left[ \frac{T_{PMS}-T_{exact}}{T_{exact}} \right] \ 
\times 100$, for $A=10$ and $\mu=1$ as a function of the order. 
The three sets are obtained by using the optimal value 
$\lambda_{PMS} = \sqrt{3 \ \mu A^2}/2$ (plus), 
a value $\lambda = 0.9 \ \lambda_{PMS}$ (triangle) and 
$\lambda = 1.1 \ \lambda_{PMS}$ (square).}
\label{Fig_1}
\end{center}
\end{figure}

Given that it is possible to calculate analytically all the integrals $\ell_{-1/2 n}$, we 
are able to obtain the exact series representation:
\begin{equation}
T_\delta = \sum_{n=0}^\infty  \delta^n 
\frac{(-1)^n\pi(2 n-1)!!}{2^{2 n-1}n!\sqrt{1+\lambda^2}}
\left( \frac{A^2  \mu - 2  \lambda^2}{1+\lambda^2} \right)^n \
_2F_1\left(\frac{1}{2},-n,1,\frac{A^2\mu}{2  \lambda^2-A^2  \mu}\right) ,
\label{eq_2_6}
\end{equation}
where $_2F_1$ is the hypergeometric function. Since eq.~(\ref{eq_2_6}) is 
essentially a power series, it converges exponentially to the exact result, which 
is precisely what we observe in  Figure~\ref{Fig_1}, where  we plot the error 
$\Xi \equiv \left[ \frac{T_{PMS}-T_{exact}}{T_{exact}} \right] \ \times 100$ for 
three different values of the parameter $\lambda$ as a function of the order in the expansion. 
$T_{exact}$ is the exact period of the Duffing oscillator which can be expressed in terms of elliptic functions. 
Corresponding to the optimal value of the parameter,  $\lambda_{PMS} = \sqrt{3 \ \mu} A/2$, the rate of 
convergence is maximal.
To the best of my knowledge eq.~(\ref{eq_2_6}) corresponding to $\lambda_{PMS}$ provides the fastest converging 
series representation of the period of the Duffing oscillator.

We consider now the nonlinear pendulum, whose potential is given by
$V(\theta) = 1 - \cos\theta$.
By choosing the interpolating potential to be $V_0(\theta) = \frac{1+\lambda^2}{2} 
\theta^2$ we obtain
\begin{equation}
\Delta(\theta) = - \frac{2}{1+\lambda^2} 
 \frac{\cos\Theta - \cos\theta}{\Theta^2-\theta^2} -1 ,
\label{eq_2_7}
\end{equation}
where $\Theta$ is the amplitude of the oscillations.
To first order our formula yields
\begin{equation}
T_\delta = \frac{2\pi}{\sqrt{1+\lambda^2}} \Big( 1 + \frac{\delta}{2} \Big) - 
\frac{2  \pi  \delta}{(1+\lambda^2)^{3/2}}\frac{J_1(\Theta)}{\Theta},
\label{eq_2_8}
\end{equation}
where $J_1$ is the Bessel function of the first kind of order 1.
The optimal value of $\lambda$ in this case is given by
\begin{equation}
\lambda_{PMS} = \sqrt{\frac{2J_1(\Theta)}{\Theta}-1}
\label{eq_2_9}
\end{equation}
and the period to first order is then
\begin{equation}
T_{PMS} = \pi\sqrt{\frac{2 \ \Theta}{J_1(\Theta)}}.
\label{eq_2_10}
\end{equation}

Despite its simplicity eq.~(\ref{eq_2_10}) provides an excellent approximation to the exact period over
a wide range of amplitudes.

\subsection{General Relativity}

We now apply our expansion to two problems in General Relativity: 
the calculation of the deflection of the light by the Sun and the
calculation of the precession of a planet orbiting around the Sun. 
We use the notation of Weinberg~\cite{Weinberg}:
\begin{equation}
B(r) = A^{-1}(r) = 1 - \frac{2 G  M}{r}.
\label{eq_2_11}
\end{equation}

The angle of deflection of the light by the Sun is given by the expression
\begin{equation}
\Delta \phi = 2  \int_{r_0}^\infty  \sqrt{A(r)}  
\left[ \left(\frac{r}{r_0}\right)^2  \frac{B(r_0)}{B(r)} -1\right]^{-1/2} 
\frac{dr}{r} - \pi 
\label{eq_2_12}
\end{equation}
where $r_0$ is the closest approach.

With the change of variable $z = 1/r$ we obtain
\begin{equation}
\Delta \phi = 2 \ r_0^{3/2}  \int_0^{1/r_0} 
 \frac{dz}{\sqrt{r_0 - r_0^3 z^2 - 2GM (1- r_0^3  z^3)}} - \pi,
\label{eq_2_13}
\end{equation}
which is exactly in the form of eq.~(\ref{eq_1_1}). We introduce the
potential $V_0(z) = (r_0^3 + \lambda^2) \ z^2$ and obtain
\begin{equation}
\Delta(z) = -\frac{\lambda^2 (z^2 -\frac{1}{r_0^2}) - 2  G M \ (1 + r_0^3  z^3)}{(r_0^3 + \lambda^2)
\left( z^2 -\frac{1}{r_0^2} \right)}.
\label{eq_2_14}
\end{equation}

By performing the standard steps which are required by our method we obtain the optimal deflection angle
to first order to be:
\begin{equation}
\Delta\phi_{PMS} = - \pi + \sqrt{\frac{\pi}{1 - \displaystyle
\frac{8  G  M}{r_0  \pi}}}.
\label{eq_2_15}
\end{equation}
corresponding to the optimal $\lambda$:
\begin{equation}
\lambda_{PMS}^2 = - \frac{8  G  M r_0^2}{\pi} .
\label{eq_2_16}
\end{equation}

The surface corresponding to the closest approach for which $\Delta\phi$ 
diverges is known as {\sl photon sphere} and for the Schwartzchild metric 
takes the value $r_0 = 3 G M$. 
It is remarkable that eq.~(\ref{eq_2_16}), despite its simplicity, 
is able to predict a slightly smaller {\sl photon sphere}, corresponding to 
$r_0 = 8 G M/\pi$. This feature is missed  completely in a perturbative approach.

\begin{figure}
\begin{center}
\includegraphics[width=9cm]{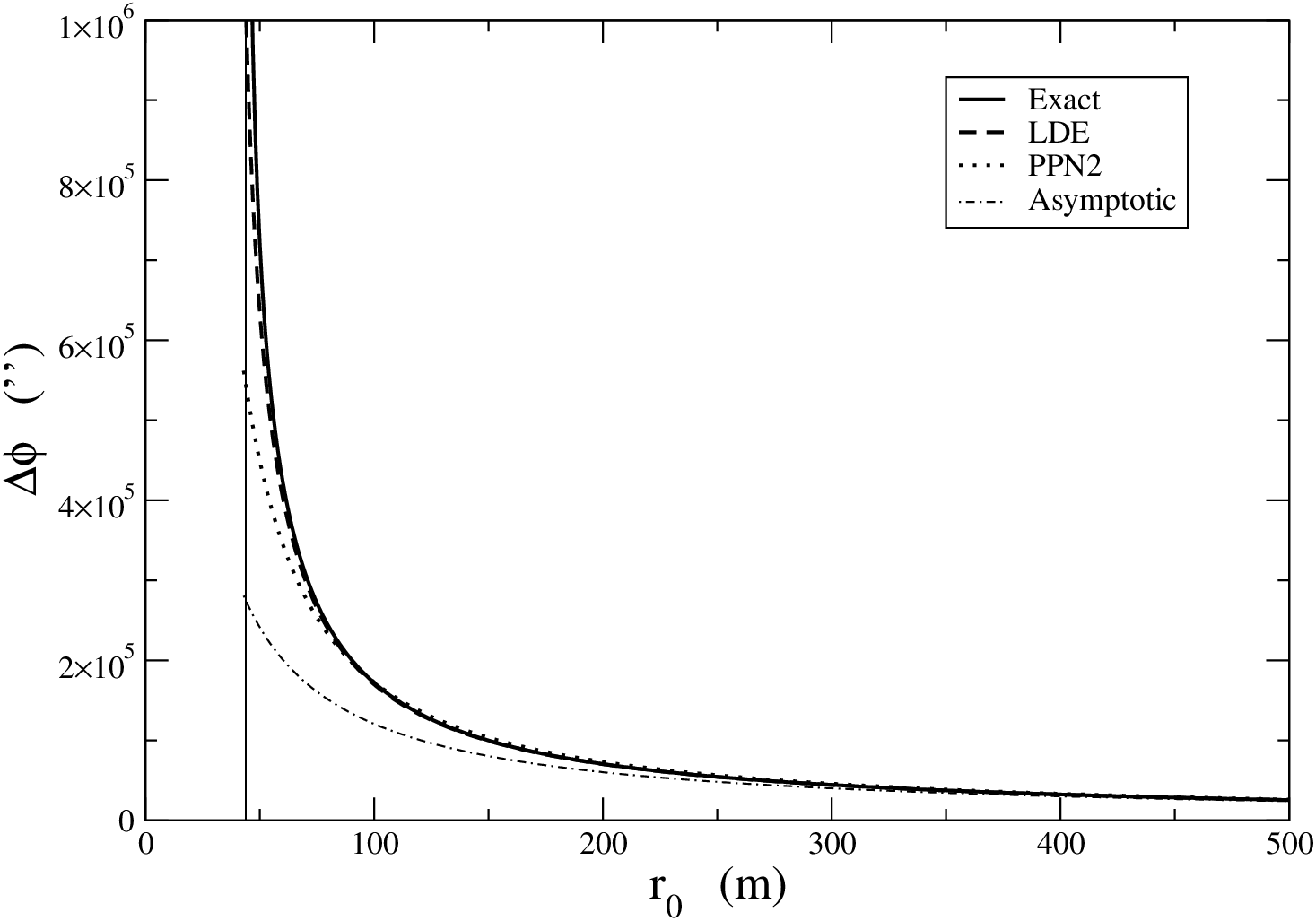}
\caption{Deflection angle of light obtained assuming 
$G/c^2 = 7.425 \times 10^{-30} \ m/kg$ and $M = 1.97 \times 10^{30} \ kg$ as 
function of the closest approach $r_0$. The solid line is the exact (numerical) 
result, the dashed line is obtained with eq.~(\ref{eq_2_16}), the dotted 
line is the post-post-Newtonian result of \cite{Epstein}, the dot-dashed line 
is the asymptotic result ($r_0 \rightarrow \infty$).
The vertical line marks the location of the photon sphere, where the deflection 
angle diverges.}
\label{Fig_2}
\end{center}
\end{figure}

\begin{figure}
\begin{center}
\includegraphics[width=9cm]{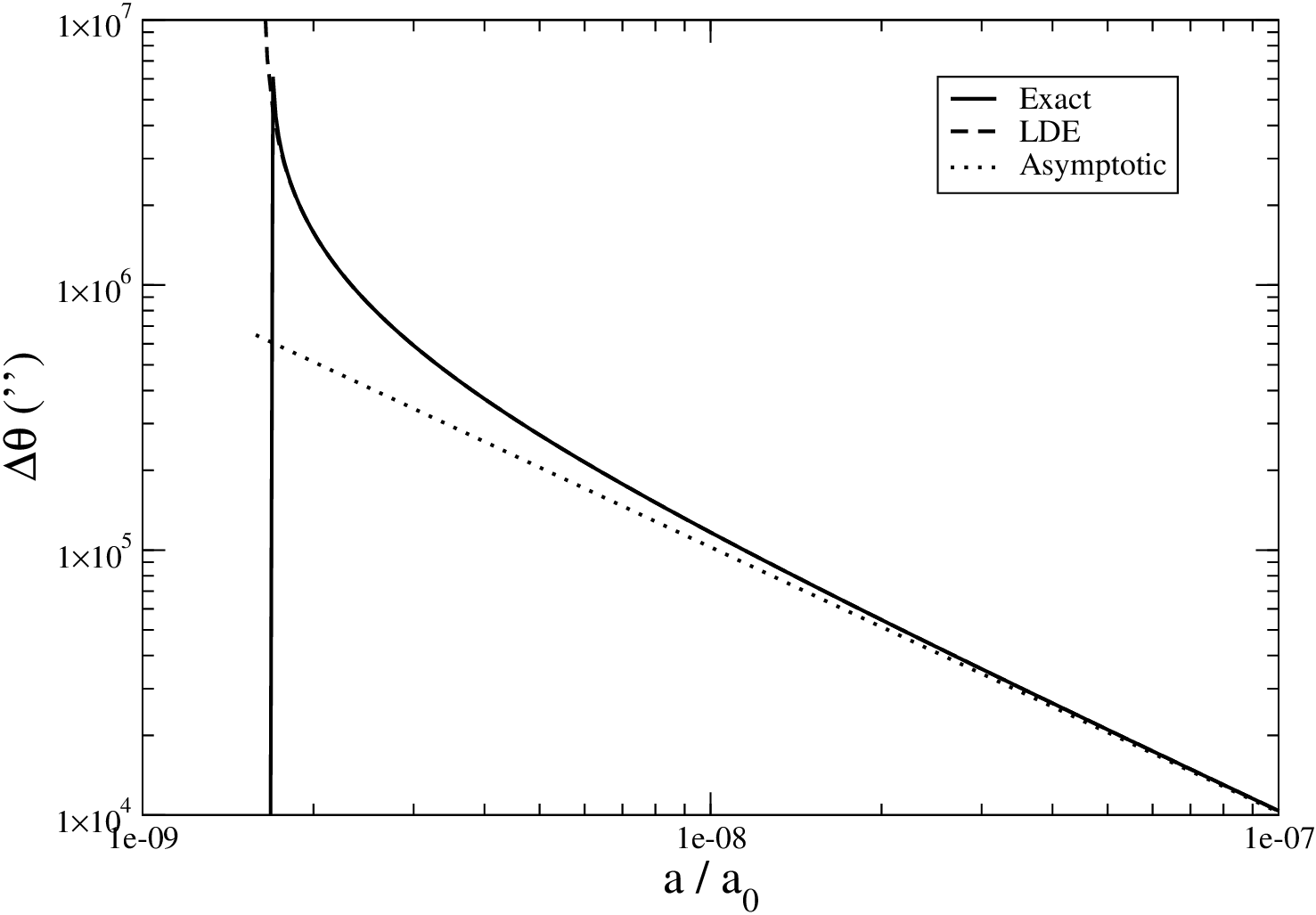}
\caption{Precession of the orbit of a planet assuming the values 
$M = 1.97 \times 10^{30} \ kg$, $G/c^2 = 7.425 \times 10^{-30} \ m/kg$ and
$\varepsilon = 0.2506$ (eccentricity). The scale of reference $a_0$ is taken to 
be the semimajor axis of Mercury's orbit ($a_0 = 5.971 \times 10^{10} \ 
m$). The solid line is the exact result, the dashed line is the result of 
eq.~(\ref{eq_2_19}) and the dotted line is the 
leading term in the perturbative expansion.}
\label{Fig_3}
\end{center}
\end{figure}

In Figure~\ref{Fig_2} we compare eq.~(\ref{eq_2_16})
with the exact numerical result, the post-post-Newtonian (PPN) result 
of \cite{Epstein} and with the asymptotic result for very small
values of $r_0$ (close to the photon sphere). We assume $M$ to correspond to the
physical mass of the Sun and $G$ to the physical value of the gravitational constant.
This corresponds to a strongly nonperturbative regime, where the gravitational force 
is extremely intense. The reader can judge the quality of our approximation.

We now consider the problem of calculating the precession of the perihelion of a
planet orbiting around the Sun. The angular precession is given by~\cite{Weinberg}
\begin{equation}
\Delta \theta = -2  \pi + 2  \int_{r_-}^{r_+}
\frac{\sqrt{A(r)}\ dr}{r^2\sqrt{\displaystyle\frac{1}{J^2 B(r)} - \frac{E}{J^2} - 
\frac{1}{r^2}}}
\label{eq_2_17}
\end{equation}
where
$E = \left(\frac{r_+^2}{B(r_+)}-\frac{r_-^2}{B(r_-)}\right)/\left(r_+^2-r_-^2\right)$
and $J^2 = \left(\frac{1}{B(r_+)}-\frac{1}{B(r_-)}\right)/\left(\frac{1}{r_+^2}-\frac{1}{r_-^2}\right)$.
$r_\pm$ are the shortest (perielia) and largest (afelia) distances from the sun. 
By the change of variable $z = 1/r$ we can write eq.~(\ref{eq_2_17}) as
\begin{equation}
\Delta \theta = - 2  \int_{z_-}^{z_+}  \frac{1}{\sqrt{(z_+-z) (z-z_-)}} \
\frac{dz}{\sqrt{(1- 2 G M (z+z_-+z_+))}} - 2  \pi,
\label{eq_2_18}
\end{equation}
where $z_\pm \equiv 1/r_\mp$. Once again the integral has the form required by our method. 
One obtains
\begin{equation}
\Delta \theta = 2 \pi \ \\  \left[
\frac{\pi(3 G^2 L M^2+ a ( -4 L^2 + 48 G L M - 147 G^2 M^2 ))}%
{4a (L - 6GM)^2 \sqrt{1 - \displaystyle\frac{6\ G\ M}{L}}} - 1 \right],
\label{eq_2_19}
\end{equation}
where $a$ is the semimajor axis of the ellipse, given by $a \equiv (r_-+r_+)/2$,
and $L$ is the {\sl semilatus rectum} of the ellipse, given by $1/L = (1/r_++1/r_-)/2$.
The optimal $\lambda$ is 
\begin{equation}
\lambda_{PMS} =  \sqrt{\frac{6 G M}{L}} .
\label{eq_2_20}
\end{equation}

In Figure~\ref{Fig_3} we plot the precession of the orbit calculated 
through the exact formula (solid line), through eq.~(\ref{eq_2_19}) (dashed line) and 
through the leading order result $\Delta\theta_0 = \frac{6\pi GM}{L}$ (dotted line)~\cite{Weinberg}  . Once again we find excellent 
agreement with the exact result.

\subsection{The spectrum of a quantum potential}

In \cite{AL04} the method described in this paper was applied to the calculation of the spectrum of an anharmonic potential within the WKB 
method to order $\hbar^6$. The WKB condition is 
\begin{eqnarray}
\Lambda(E) = \frac{\pi \hbar}{\sqrt{2 m}} \ (n+1/2)
\label{eq_2_21}
\end{eqnarray}
where $\Lambda(E)$ to order $O(\hbar^6)$ is given by
\begin{eqnarray}
\Lambda(E) &\equiv& {\cal J}_{1}(E) -  \frac{\hbar^2}{48 \ m} \frac{d}{dE} {\cal J}_{2}(E) + 
\frac{\hbar^4}{11520 \ m^2} \frac{d^3}{dE^3}  {\cal J}_{3}(E) \ .
\label{eq_2_22}
\end{eqnarray}

\begin{figure}
\begin{center}
\includegraphics[width=9cm]{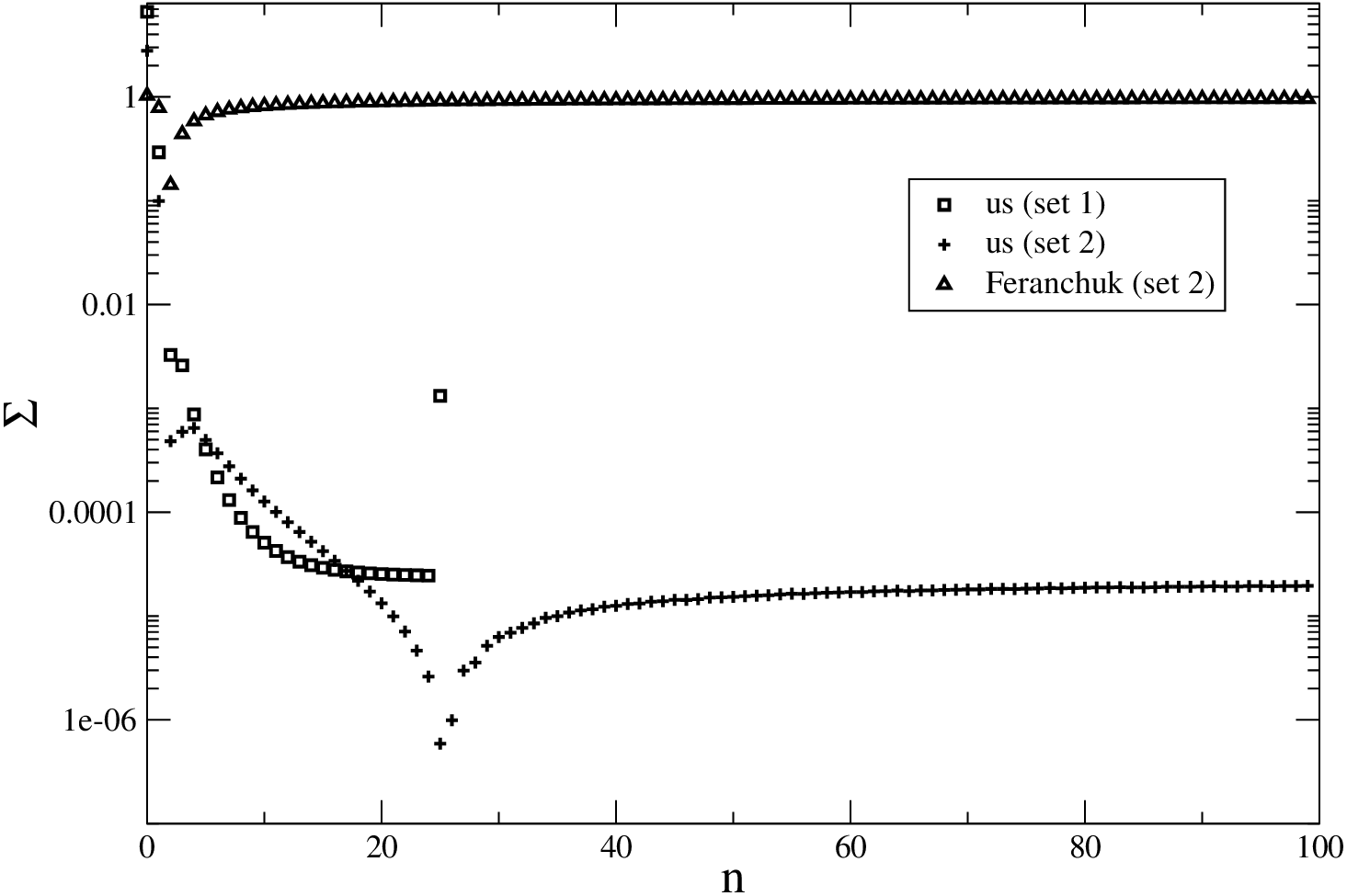}
\caption{Error over the energy of the anharmonic oscillator with  $\hbar = 1$, $m= 1/2$, $\omega= 2$ and $\mu = 8000$ (set 1) and
with $\hbar = m= \omega= 1$ and $\mu = 4$ (set 2). The boxes and the pluses have been obtained with our method, the triangles 
correspond to the error calculated using the analytical formula of \cite{Fer95}.}
\label{Fig_5}
\end{center}
\end{figure}

We have defined the integrals:
\begin{eqnarray}
\label{eq_2_23}
{\cal J}_{1}(E) &\equiv&  \int_{x_-}^{x_+} \sqrt{E-V(x)} dx \\
\label{eq_2_24}
{\cal J}_{2}(E) &\equiv&  \int_{x_-}^{x_+} \frac{V''(x)}{\sqrt{E-V(x)}} dx \\
\label{eq_2_25}
{\cal J}_{3}(E) &\equiv& \int_{x_-}^{x_+} \frac{7 \ V''(x)^2 - 5 \ V'(x) \ V'''(x)}{\sqrt{E-V(x)}} dx \ ,
\end{eqnarray}
where $x_{\pm}$ are the classical turning points. The spectrum of the potential $V(x)$ can be obtained by 
solving eq.~(\ref{eq_2_22}). The integrals appearing in the equations (\ref{eq_2_23}),  (\ref{eq_2_24}) and
 (\ref{eq_2_25}) are of the form required by eq.~(\ref{eq_1_1}). We can test the method with the 
quantum anharmonic potential $V(x) = \frac{m \omega^2 x^2}{2} + \frac{\mu x^4}{4}$: in \cite{AL04} 
it was proved that the integrals above can be analytically approximated with very high precision with our method.

By solving eq.~(\ref{eq_2_21}) once that the integrals have been approximated with our method one
obtaines an {\sl analytical} formula for the spectrum of the anharmonic oscillator~\cite{AL04}:
\begin{eqnarray}
E_n &\approx& e_1 \ \left(n+\frac{1}{2}\right)^{\frac{4}{3}} + e_2 \ \left(n+\frac{1}{2}\right)^{\frac{2}{3}} + 
e_3 + \frac{e_4}{\left( n+\frac{1}{2}\right)^{2/3}} + \dots 
\label{EQN6}
\end{eqnarray}
where the first few coefficients $e_i$ are given by
\begin{eqnarray}
e_1 &\approx& 0.867146 \ {\left( \frac{{\mu}\ {{\hbar}}^4}{m^2} \right) }^{\frac{1}{3}} \\
e_2 &\approx& 0.42551  \left( \frac{\hbar \ m}{\sqrt{\mu}} \right)^{2/3} \ \omega^2 \\
e_3 &\approx& -0.0466914 \ \frac{m^2 \ \omega^4}{\mu} \\
e_4 &\approx& 0.030669 \left( \frac{\mu \hbar^4}{m^2} \right)^{1/3} + 0.00424238 \  
\left( \frac{m^{10} \omega^6}{\mu^5 \hbar^2}\right)^{1/3} \ .
\end{eqnarray}

In Fig.~\ref{Fig_5} we display the error over the energy defined as 
$\Sigma \equiv \left| \frac{E_n^{(approx)}-E_n^{(exact)}}{E_n^{(exact)}}\right| \times 100$ as a function of the quantum 
number $n$. The boxes have been obtained using our formula eq.~(\ref{EQN6}) and assuming $\hbar = 1$, $m= 1/2$, $\omega= 2$ and $\mu = 8000$. 
In this case $E_n^{(exact)}$ are the energies of the anharmonic oscillator calculated with high precision in last column of Table III of 
\cite{mei97}. The jump corresponding to $n=25$ is due to the low precision of the last value of Table III of \cite{mei97}. 
The pluses and the triangles have been obtained using  our formula eq.~(\ref{EQN6}) (pluses) and eq.~(1.34) of \cite{Fer95} (triangles)
and assuming $\hbar = m= \omega= 1$ and $\mu = 4$. In this case $E_n^{(exact)}$ are the energies of the anharmonic oscillator numerically 
calculated through a fortran code. We can easily appreciate that our formula provides an approximation which is several orders of magnitude better
than the one of eq.~(1.34) of \cite{Fer95}. 
We also notice that the formula  of \cite{Fer95} yields a quite different asymptotic expansion in the limit of $n \gg 1$.
We are not aware of expressions for the spectrum of the anharmonic oscillator similar to the one given by eq.~(\ref{EQN6}).

\subsection{The Riemann zeta function}

The method outlined above can be applied also to the calculation of the Riemann zeta function~\cite{Am04}.
We  consider the integral representation
\begin{eqnarray}
\zeta(n) &=& \frac{(-2)^{n-1}}{2^{n-1}-1} \frac{1}{\Gamma(n)} \int_0^1 \frac{\log^{n-1} x}{1+x} dx \ .
\label{s4_1}
\end{eqnarray}

Although eq.~(\ref{s4_1}) is not of the standard form of eq.~(\ref{eq_1_2}), we can write it as:
\begin{eqnarray}
\zeta(n) &=& \frac{(-2)^{n-1}}{2^{n-1}-1} \frac{1}{\Gamma(n)} \int_0^1 \frac{1}{1+\lambda} \ 
\frac{\log^{n-1} x}{1+\frac{x-\lambda}{1+\lambda}} dx  ,
\label{s4_2}
\end{eqnarray}
where $\lambda$ is as usual an arbitrary parameter introduced by hand. In this case $\Delta(x) \equiv \frac{x-\lambda}{1+\lambda}$
and the condition $\left|\frac{x-\lambda}{1+\lambda}\right| < 1$ is fullfilled provided that $\lambda >0$; 
one can expand the denominator in powers of $\frac{x-\lambda}{1+\lambda}$ and obtain:
\begin{eqnarray}
\zeta(n) &=&  \frac{(2)^{n-1}}{2^{n-1}-1}  \sum_{k=0}^\infty \frac{1}{(1+\lambda)^{k+1}} \sum_{j=0}^k \ \frac{k!}{j! (k-j)!} 
\lambda^{k-j}  \frac{(-1)^{j} }{(1+j)^n} \ .
\label{s4_3}
\end{eqnarray}

Despite its appearance this series {\sl does not depend} upon $\lambda$, 
as long as $\lambda >0$. This means that when the sum over $k$ is truncated to a given finite order a residual
dependence upon $\lambda$ will survive: such dependence will be minimized by applying the PMS~\cite{Ste81}, i.e.
by asking that the derivative of the partial sum with respect to $\lambda$ vanish. 
To lowest order one has that $\lambda_{PMS}^{(1)} = 2^{-n}$ and the corresponding formula is found:
\begin{eqnarray}
\zeta(n) &=&  \frac{(2)^{n-1}}{2^{n-1}-1}  \sum_{k=0}^\infty \frac{1}{(1+2^{-n})^{k+1}} \sum_{j=0}^k \ 
\frac{k!}{j! (k-j)!} 2^{-n (k-j)}  \frac{(-1)^{j} }{(1+j)^n}  .
\label{s4_4}
\end{eqnarray}

\begin{figure}
\begin{center}
\includegraphics[width=9cm]{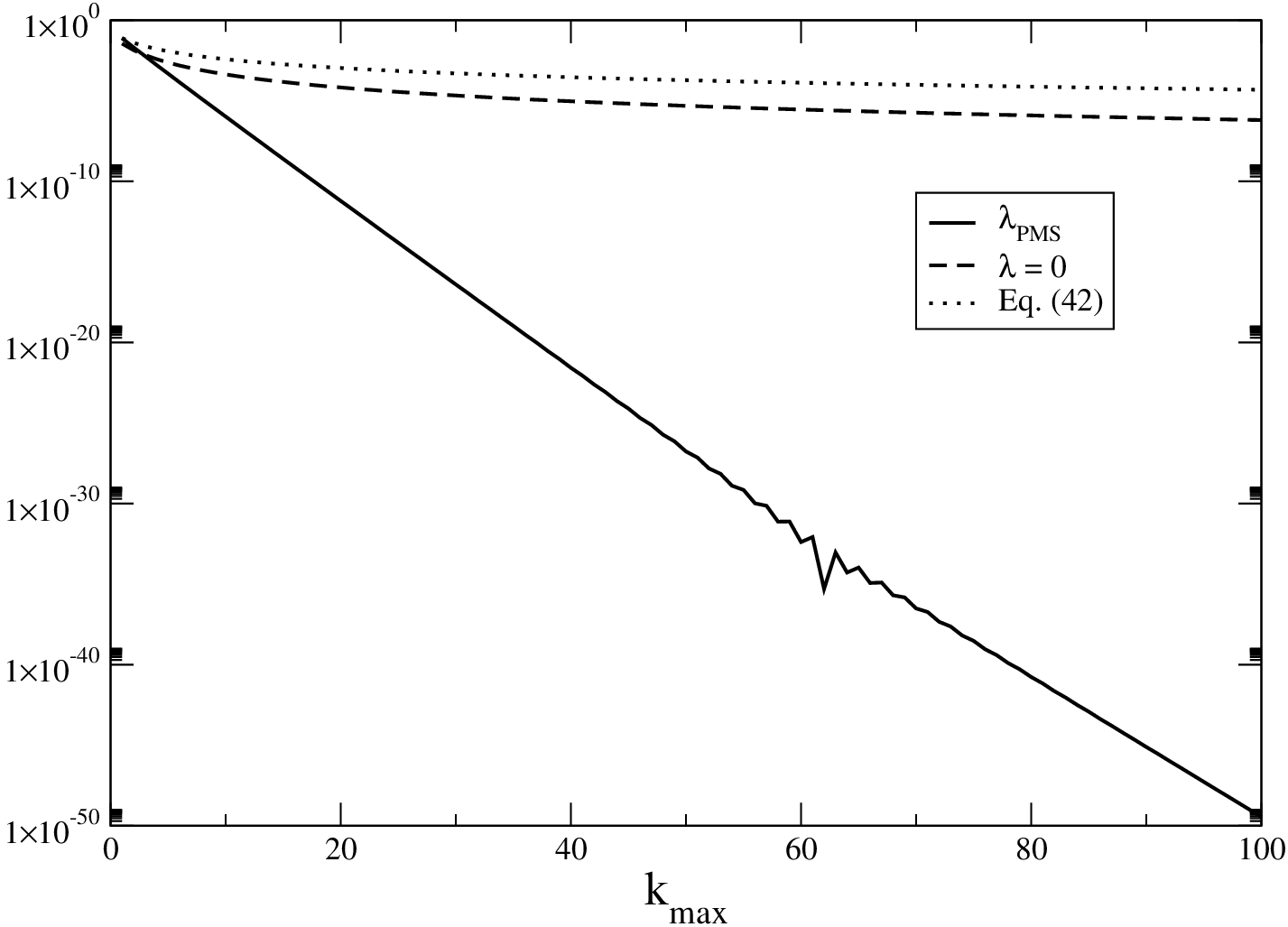}
\caption{Difference  $|\zeta_{app}(3)-\zeta(3)|$  as a function of the number of terms in the sum.\label{fig6}}
\end{center}
\end{figure}

We want to stress that eq.~(\ref{s4_4}) is still an {\sl exact} series representation of the Riemann zeta function.
This simple formula yields an excellent approximation to the zeta function even in proximity of $s=1$ where the function 
diverges. The rate of convergence of the series is greatly improved by applying the PMS to higher 
orders. In Fig.~\ref{fig6} we plot the difference $|\zeta_{app}(3)-\zeta(3)|$ using eq.~(\ref{s4_3}) with 
$\lambda = \lambda_{PMS}$ (solid line), $\lambda = 0$ (dashed line) and the series representation
\begin{eqnarray}
\zeta(3) = \sum_{n=0}^\infty \frac{1}{(n+1)^3} 
\label{s4_5}
\end{eqnarray}
which corresponds to the dotted line in the plot. This last series converges quite slowly and a huge number
of terms (of the order of $10^{25}$) is needed to obtain the same accuracy that our series  
with $\lambda_{PMS}$ reaches with just $10^2$ terms.
We notice that a special case of eq.~(\ref{s4_3}), corresponding to $\lambda = 1$, was already known in the 
literature~\cite{Kno30}. 

We can extend eq.~(\ref{s4_3}) to the critical line, $s = \frac{1}{2} + i \tau$, and write
\begin{eqnarray}
\zeta\left(\frac{1}{2} + i \tau\right) &=& \frac{(2)^{-\frac{1}{2} + i \tau}}{2^{-\frac{1}{2} + i \tau}-1}  
\sum_{k=0}^\infty \frac{1}{(1+\lambda)^{k+1}} 
\sum_{j=0}^k \ \frac{k!}{j! (k-j)!}   \frac{(-1)^{j} \lambda^{k-j} }{(1+j)^{\frac{1}{2} + i \tau}} 
\label{eq_3}
\end{eqnarray}

\begin{figure}
\begin{center}
\includegraphics[width=9cm]{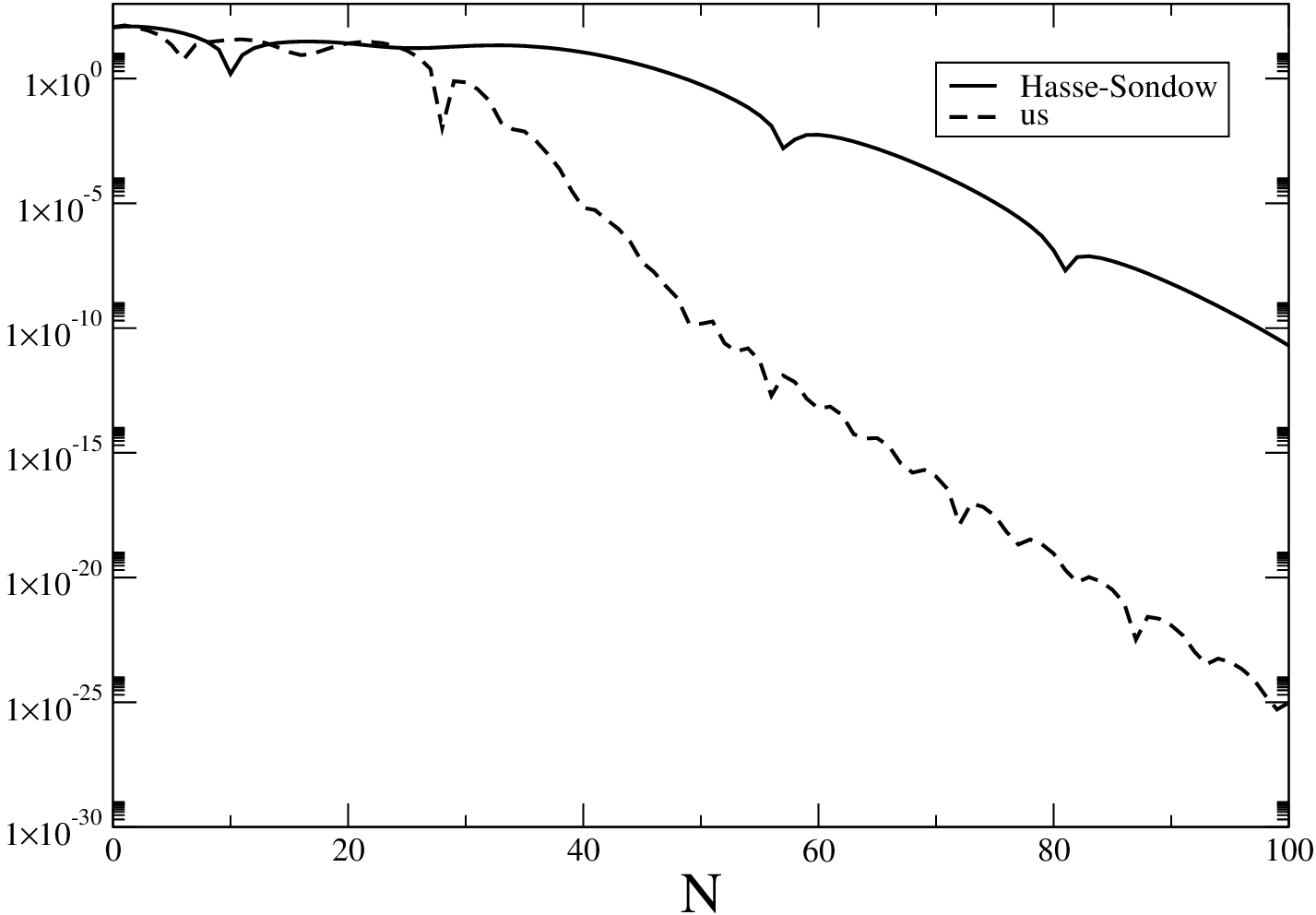}
\caption{$\Xi \equiv Re \left[\frac{\zeta^{(N)}\left(\frac{1}{2} + i \tau\right) -\zeta\left(\frac{1}{2} + i \tau\right)}{\zeta\left(\frac{1}{2} + i \tau\right)}\right] \times 100$, as a function of the number of terms considered in the sum of eq.~(\ref{eq_3}). The dashed curve is
obtained using $\lambda = 0.3$.
\label{fig7}}
\end{center}
\end{figure}

In Fig.~\ref{fig7} we have plotted the error (in percent) over the real part of the zeta function, i.e.
$\Xi \equiv Re \left[\frac{\zeta^{(N)}\left(\frac{1}{2} + i \tau\right) -\zeta\left(\frac{1}{2} + i \tau\right)}{\zeta\left(\frac{1}{2} + i \tau\right)}\right] \times 100$, as a function of the number of terms considered in the sum of eq.~(\ref{eq_3}). We use $\tau = 50$. 
The solid line corresponds to the formula of \cite{Kno30}, whereas the dashed line corresponds to using our formula, 
eq.~(\ref{eq_3}), with $\lambda = 0.3$. 

\begin{figure}
\begin{center}
\includegraphics[width=9cm]{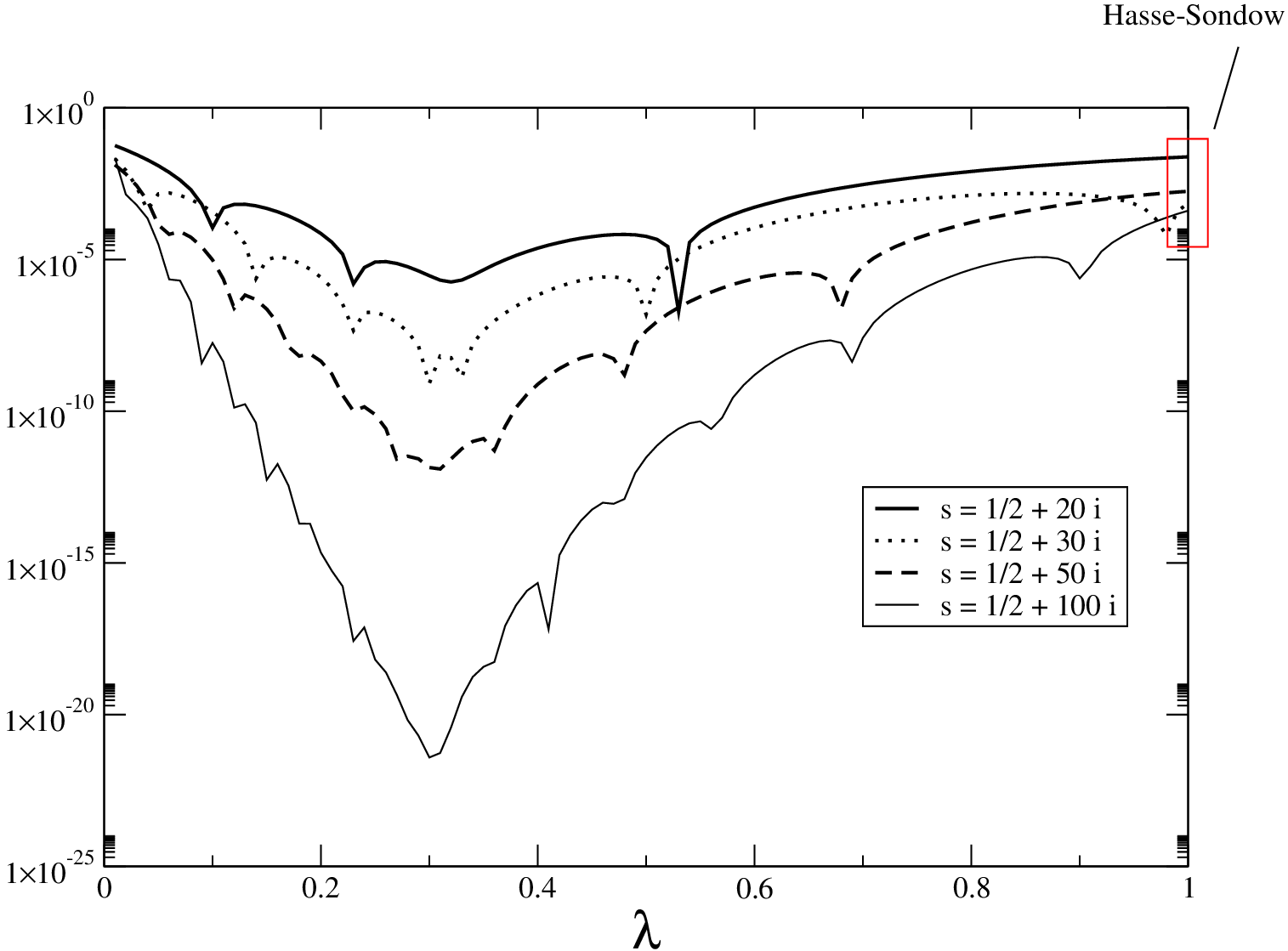}
\caption{The error obtained using the partial sum of eq.~(\ref{eq_3}) for $s = \frac{1}{2} + i \tau$ and
taking the first $\tau$ terms in the sum.\label{fig8}}
\end{center}
\end{figure}

In Fig.~\ref{fig8} we plot the difference $\zeta^{(\tau)}\left(\frac{1}{2} + i \tau\right) -\zeta\left(\frac{1}{2} + i \tau\right)$ as a 
function of $\lambda$ for different values of $\tau$. In this case the series is limited to the first $\tau$ terms. The optimal value of 
$\lambda$ is found close to $\lambda = 0.3$.

The two figures prove that our expansion is greatly superior to the one of \cite{Kno30}.

\section{Conclusions}
\label{conclu}

In this paper we have reviewed a method which allows to estimate a certain class of integrals with arbitrary
precision. The method is based on  the Linear Delta Expansion, i.e. on the powerful idea that a certain 
(unsoluble) problem can be interpolated with a soluble one, depending upon an arbitrary parameter and then performing 
a perturbative expansion. The principle of minimal sensitivity allows one to obtain results which
converge quite rapidly to the correct results. It is a common occurrence in calculations based on Variational Perturbation
Theory, like the present one, that the solution to the PMS equation to high orders cannot be performed analytically. 
Here, however, we do not face this problem, since we have proved that the method converges in a whole region in the parameter space: 
the convergence of the expansion is granted as long as the parameter falls in that region.

Although in this paper we have examined a good number of applications 
of this method to problems both in Physics and Mathematics, we feel that it can be used, with minor modifications, 
in dealing with many other problems. An extension of the method in this direction is currently in progress.

\bigskip

\bigskip

The author acknowledges support of Conacyt grant no. C01-40633/A-1. He also thanks the organizing comitee 
of the {\sl Dynamical Systems, Control and Applications} (DySCA) meeting for the kind invitation to participate
to the workshop.

%\begin{thebibliography}
%\bibitem{N81}  A. H. Nayfeh, Introduction to Perturbation Techniques (John Wiley and Sons, New York, 1981).
%\end{thebibliography}


\begin{thebibliography}{99}
  
\bibitem{AS04}  P. Amore and R. A. S\'{a}enz, The Period of a Classical Oscillator, ArXiv:[math-ph/0405030].

\bibitem{AAFS04}  P. Amore, A. Aranda, F. M. Fern\'{a}ndez, and R.
S\'{a}enz, Systematic Perturbation of Integrals with Applications to
Physics,  ArXiv:[math-ph/0407014].

\bibitem{AF04} P. Amore and  F. M. Fern\'{a}ndez, Exact and approximate expressions for the period of anharmonic oscillators, ArXiv:[math-ph/0409034].

\bibitem{AL04} P. Amore and J. Lopez, The spectrum of a quantum potential,  ArXiv:[quant-ph/0405090].

\bibitem{Am04} P. Amore, Convergence acceleration of series through a variational approach,  ArXiv:[math-ph/0408036].

\bibitem{lde}            A. Okopi\'nska, Phys.\ Rev.\ D {\bf 35}, 1835 (1987); A. Duncan and M. Moshe, 
                         Phys.\ Lett.\ B {\bf 215}, 352 (1988)

\bibitem{Ste81}          P. M. Stevenson,
                         Phys. Rev. D {\bf 23}, 2916 (1981).

\bibitem{AA1:03}         P. Amore and A. Aranda,  Phys. Lett. A {\bf 316} 218

\bibitem{Weinberg}       S. Weinberg, Gravitation and cosmology, J.Wiley and Sons, 1972

\bibitem{Epstein}        R. Epstein and I. Shapiro, Phys.\ Rev.\ D {\bf 22}, 2947 (1980);
                         E. Fischbach and B. Freeman,  Phys.\ Rev.\ D {\bf 22}, 2950 (1980)

\bibitem{mei97}          H. Meissner and O. Steinborn,
                         Phys. Rev. A {\bf 56}, 1189 (1997).

\bibitem{Fer95}          I.D. Feranchuk, L.I. Komarov, I.V. Nichipor  and A.P. Ulyanenkov, Annals of Physics {\bf 238} 370 (1995)

\bibitem{Kno30} K. Knopp,  "4th Example: The Riemann -Function." Theory of Functions Parts I and II, Two Volumes Bound as One, 
Part II. New York: Dover, pp. 51-57, 1996; H. Hasse, "Ein Summierungsverfahren für die Riemannsche Zeta-Reihe." Math. Z. 32, 458-464, 1930;
J. Sondow, "Analytic Continuation of Riemann's Zeta Function and Values at Negative Integers via Euler's Transformation of Series." Proc. Amer. Math. Soc. 120, 421-424, 1994. 


\end{thebibliography}
\end{document}